\documentclass[12pt]{article}
\usepackage{amsmath,amssymb,euscript ,yfonts,psfrag,latexsym,dsfont,graphicx,bbm,color,amstext,wasysym,subfig,parskip}
\usepackage{flushend}
\graphicspath{{./},{../figures/}}

\begin{document}
\newtheorem{thm}{Theorem}
\newtheorem{cor}[thm]{Corollary}
\newtheorem{conj}[thm]{Conjecture}
\newtheorem{lemma}[thm]{Lemma}
\newtheorem{prop}{Proposition}
\newtheorem{problem}[thm]{Problem}
\newtheorem{remark}{Remark}
\newtheorem{defn}[thm]{Definition}
\newtheorem{ex}[thm]{Example}

\newcommand{\N}{{n}}
\newcommand{\mR}{{\mathbb R}}
\newcommand{\D}{{\mathbb D}}
\newcommand{\E}{{\mathbb E}}
\newcommand{\cN}{{\mathcal N}}
\newcommand{\mcR}{{\mathcal R}}
\newcommand{\mcS}{{\mathcal S}}
\newcommand{\cS}{{\mathcal S}}
\newcommand{\cC}{{\mathcal C}}
\newcommand{\diag}{\operatorname{diag}}
\newcommand{\tr}{\operatorname{trace}}
\newcommand{\f}{{\mathfrak f}}
\newcommand{\g}{{\mathfrak g}}
\newcommand{\range}{\mcR}\newcommand{\trace}{\operatorname{trace}}
\newcommand{\zero}{{0}}

\newcommand{\ignore}[1]{}

\def\spacingset#1{\def\baselinestretch{#1}\small\normalsize}
\setlength{\parskip}{10pt}
\setlength{\parindent}{20pt}
\spacingset{1}
\newcommand{\mike}{}
\definecolor{grey}{rgb}{0.6,0.6,0.6}
\definecolor{lightgray}{rgb}{0.97,.99,0.99}

\title{Optimal steering of inertial particles\\ {\mike diffusing anisotropically with losses}}

\author{Yongxin Chen, Tryphon Georgiou and Michele Pavon
\thanks{Y.\ Chen and T.T.\ Georgiou are with the Department of Electrical and Computer Engineering,
University of Minnesota, Minneapolis, Minnesota MN 55455, USA, \{chen2468,tryphon\}@umn.edu, and M.\ Pavon is with the Dipartimento di Matematica,
Universit\`a di Padova, via Trieste 63, 35121 Padova, Italy, pavon@math.unipd.it}}
\markboth{\today}{}

\maketitle

{\mike
\begin{abstract}
Exploiting a fluid dynamic formulation for which a probabilistic counterpart might not be available, we extend the theory of Schr\"odinger bridges to  the case of inertial particles with losses and general, possibly singular diffusion coefficient. We find that, as for the case of constant diffusion coefficient matrix, the optimal control law is obtained by solving a system of two p.d.e.'s involving adjoint operators and coupled through their boundary values. In the linear case with quadratic loss function, the system turns into two matrix Riccati equations with coupled split boundary conditions. An alternative formulation of the control problem as a semidefinite programming problem allows computation of suboptimal solutions. This is illustrated in one example of inertial particles subject to a constant rate killing.
\end{abstract}}

\section{Introduction}
In 1931/1932, Erwin Schr\"odinger \cite{Schrodinger1}, \cite[Section VII]{Schrodinger2} posed the following problem: a large number of i.i.d.\ Brownian particles in $\mR^n$ is observed at an initial time $t=0$ to have an empirical distributions $\rho_0(x)$, and at a final time $t=T$ an empirical distribution $\rho_T(x)$. Assuming that the final distribution differs from the one dictated by the law of large numbers, Schr\"odinger sought the most likely intermediate empirical distribution for the particle trajectories. He computed that this in fact {\mike has one-time density} of the form
\[\rho(x,t)=\varphi(x,t)\hat\varphi(x,t)\]
where $\varphi_t$ and $\hat\varphi$ are suitable harmonic and co-harmonic functions; i.e., they satisfy
\begin{subequations}\label{eq:Schrodinger}
\begin{eqnarray}
\label{eq:Schrodinger1}
\frac{\partial\varphi(x,t)}{\partial t}+\frac{1}{2}\sum_{i,j=1}^\N a_{ij}\frac{\partial^2(\hat\varphi(x,t))}{\partial x_i\partial x_j}&=&0,\\
\label{eq:Schrodinger2}
\frac{\partial\hat\varphi(x,t)}{\partial t}-\frac{1}{2}\sum_{i,j=1}^\N\frac{\partial^2(a_{ij}\hat\varphi(x,t))}{\partial x_i\partial x_j}&=&0, \mbox{ with}\\
\label{eq:Schrodinger3}
&&\hspace*{-6cm}\varphi(x,0)\hat\varphi(x,0)=\rho_0(x)\mbox{ and }\varphi(x,T)\hat\varphi(x,T)=\rho_T(x).
\end{eqnarray}
\end{subequations}
As usual, {\mike $a=(a_{ij})$ represents a constant, positive definite} diffusion coefficient (matrix);  (\ref{eq:Schrodinger1}-\ref{eq:Schrodinger3}) is known as a {\em Schr\"odinger system}. Great many insights and generalizations followed as well connections with Quantum mechanics and, specifically, with Schr\"odinger's own famous equation (see Wakolbinger \cite{W} for a historical account until 1991). The process sought by Schr\"odinger, as well as the corresponding measure on path space that forms a {\em bridge} between beginning and ending marginals, now bear his name. In modern terms, the Schr\"odinger problem seeks to minimize relative entropy (Kullback-Leibler distance) between distributions on trajectories given the initial and final marginals \cite{F2}.

Soon afterwards, it became apparent that Schr\"odinger's bridge has a reformulation as a stochastic optimal control problem, namely to seek a minimum energy control input $u(t)$ so that
the diffusion
\[
dX(t) = u(X(t),t)dt+\sigma dw(t),
\]
with $X(0)=\xi$ a.s., $\xi$ distributed according to $\rho_0(x)$ and $w(\cdot)$ a standard Wiener process, is consistent with the empirical marginal $\rho_T(x)$. {\mike Letting $a=\sigma\sigma'$,} the Schr\"odinger bridge can be constructed by solving
\begin{subequations}\label{Sproblem}
\begin{eqnarray}
\label{eq:S1}
&&\inf_{(\rho,u)}\int_{\mR^\N}\int_{0}^{T}\frac{1}{2}u(x,t)'a^{-1}u(x,t)\rho(x,t)dtdx,\\
\label{eq:S2}
&&\frac{\partial \rho}{\partial t}+\nabla\cdot(u\rho)=\frac{1}{2}\sum_{i,j=1}^\N\frac{\partial^2(a_{ij}\rho)}{\partial x_i\partial x_j},\\
\label{eq:S3}
&&\rho(0,x)=\rho_0(x), \quad \rho(
T,y)=\rho_T(y).\end{eqnarray}
\end{subequations}
In fact, the optimizing control turns out to be
\[
u^*(x,t)=a\nabla \log\varphi(x,t)
\]
where $\varphi(x,t)$ is the {\mike space-time} harmonic function in the solution of the Schr\"odinger system. {\mike The function $\varphi$ is connected to  the {\em neutron importance function} which has an important role in perturbation theory and reactor dynamic calculations \cite{FSA}.} The optimal control interpretation relates directly to the property that the Schr\"odinger bridge represents the law which is closest to the prior measure in the sense of relative entropy; this follows from Girsanov's theory (\cite{F2,DP,PW}, see also \cite{W} and the references therein).

Interestingly, the connection between Schr\"odinger bridges and stochastic optimal control seems to have only focused on the case of non-degenerate diffusions, {\mike possibly with creation and killing \cite{DGW,W,DPP}, }in which the constant diffusion matrix $a$ is nonsingular (to some degree necessitated by the Girsanov theory).
Thus, for instance, models of inertial particles driven by stochastic forces do not fall in this category. The authors were initially motivated by this precise problem, to steer {\em inertial} particles, and by the possibility of ``cooling'' oscillators via active feedback for high resolution measuring instruments \cite{Vinante,Munakata}.
In both of these applications,  where the stochastic excitation impacts only certain direction of the state vector, i.e., where $a$ is singular, it is still possible to construct generalized Schr\"odinger bridges for (degenerate) diffusion processes and directly connect to an optimal control problem in a similar fashion. This is done in \cite{CGP,CGP2} focusing on linear dynamics and in \cite{CGP3} in a more general context of nonlinear diffusions.

Herein, following \cite{CGP,CGP2},  we {\mike pursue yet another generalization of great practical significance to general diffusion processes.  Anisotropic diffusions are important, for instance, in image processing and computer vision \cite{PM,G}.  Notice that for general diffusion processes where $\sigma=\sigma(x,t)$ viz. $a=a(x,t)$ depends, besides on time,  on the spatial variables, a probabilistic problem might not exist. Indeed, for different drift terms, the corresponding distributions on the trajectories may be mutually singular and relative entropy be always infinite. An exception is when  the control enters through the ``same channel" as the noise \cite[(5.3)]{Fischer}, \cite{CGP} or the diffusion coefficient is uniformly bounded, nonsingular and bounded away from zero \cite[p.305]{KS}. The fluid dynamic formulation (\ref{eq:S1}-\ref{eq:S3}), which resembles the celebrated Benamou-Brenier formulation of the optimal transport problem \cite{BB,Vil,AGS}, however, does make sense also in the case of a general diffusion coefficient.}
More specifically we consider
a cloud of particles with density $\rho(x,t)$, $x\in\mR^\N$, which evolves according to the transport-diffusion equation
\begin{equation}\label{Fokker-Planck}
\frac{\partial \rho}{\partial t}+\nabla\cdot(f(x,t)\rho)+V(x,t)\rho=\frac{1}{2}\sum_{i,j=1}^\N\frac{\partial^2(a_{ij}(x,t)\rho)}{\partial x_i\partial x_j},
\end{equation}
{\mike with  $\rho(x,0)=\rho_0$  a probability density.} In departure from prior works on connections to Feynman-Kac \cite{W,PW} and in accordance with our aim to be able to model inertial particles, we assume that the matrix $a(x,t)=[a_{ij}(x,t)]_{i,j=1}^{\N}$ is positive semidefinite of constant rank on all of $\mR^\N\times [\zero,T]$ with
\[
a_{ij}(x,t)=\sum_k\sigma_{ik}(x,t)\sigma_{kj}(x,t)
\]
for a matrix $\sigma(x,t)=[\sigma_{ik}(x,t)]\in\mR^{\N\times m}$ of constant rank $m\leq \N$. The presence of $V(x,t)\ge 0$ allows for the possibility of loss of mass, so that the integral of $\rho(x,t)$ over $\mR^\N$ is not necessarily constant. This flexibility allows modeling the situation where particles, obeying
\begin{equation}\label{eq:nonlinearSDE}
dX(t)=f(X(t),t)dt+\sigma(X(t),t)dw(t),
\end{equation}
are absorbed at some rate by the medium in which they travel or, if the sign of $V$ is negative, created out of this same medium \cite[p.272]{KT}. {\mike We assume here and throughout the paper that $f$ and $\sigma$ are smooth and that the operator
$$L=\sum_{i,j=1}^na_{ij}(x,t)\partial_{x_i}\partial_{x_j}+\sum_{j=1}^nf_j(x,t)\partial_{x_j}-\partial_t
$$
satisfies H\"{o}rmander's condition \cite{H,OR} and is therefore {\em hypoelliptic}. Hypoelliptic diffusions occur in many branches of science: Ornstein-Uhlenbeck stochastic oscillators, Nyquist-Johnson circuits with noisy resistors, in image reconstruction based on Petitot's model of neurogeometry of vision \cite{BCGR}, etc. The ``reweighing" the original measure of the Markov process (\ref{eq:nonlinearSDE}) when $V$ is unbounded is a delicate issue and can be accomplished via the Nagasawa transformation, see \cite[Section 8B]{W} for the details.}
We suppose that (\ref{Fokker-Planck}) represents a {\em prior} evolution and that at some point $T>0$ we measure an empirical probability density $\rho_T(x)\neq \rho(x,T)$ as dictated by \eqref{Fokker-Planck}. Thus, the model (\ref{Fokker-Planck}) is not consistent with the estimated end-point empirical distribution. However, we have reasons to believe that the actual evolution must have been close to the nominal one and that only the actual drift {\mike field} may be different and equal to  \[\tilde f(x,t):=f(x,t)+\sigma(x,t) u(x,t).\]
Notice that the control variables, which may be fewer than $n$, act through the same channels of the diffusive part. The assumption that stochastic excitation and control enter through the same ``channels'' is natural in certain applications as explained and treated in \cite{CGP} for linear diffusions. The case were these channels may differ is considered in~\cite{CGP2}.

{\mike The paper is organized as follows: The basic theory is outlined in Section \ref{sec:generalizedS} where we derive a generalized Schr\"{o}dinger system for the optimal control law. We then specialize to the case of linear dynamics with constant diffusion coefficient} and quadratic potential $V$ in Section \ref{sec:linearS} and derive a system of coupled Riccati equations that correspond to the Schr\"odinger system. The theory, even in the absence of a ``killing'' potential $V,$ falls outside the setting of ``linear-quadratic regulator theory'' {\mike where only one matrix Riccati equation occurs with a specified boundary value. Moreover, here }the coupling between the two differential Riccati equations through their split boundary conditions is non-standard, and the solutions to the Riccati equations are sign-indefinite in general \cite{CGP}. Then, in Section \ref{sec:numerics} we outline a numerical scheme to obtain suboptimal solutions for the corresponding stochastic control problem. Finally, in Section \ref{sec:example} we conclude with a numerical example.

\section{A generalized Schr\"odinger system}\label{sec:generalizedS}

{\mike Taking (\ref{Fokker-Planck}) as a reference evolution and given the terminal probability density $\rho_T$, we are led to} consider the problem
\begin{subequations}\label{FDproblem}
\begin{eqnarray}
\label{FD1}
&&\inf_{(\tilde{\rho},\tilde{u})}\int_{\mR^\N}\int_{\zero}^{T}\left[\frac{1}{2}\|u\|^2+V(x,t)\right]\tilde{\rho}(x,t)dtdx,\\
&&\frac{\partial \tilde{\rho}}{\partial t}+\nabla\cdot((f+\sigma u)\tilde{\rho})=\frac{1}{2}\sum_{i,j=1}^\N\frac{\partial^2\left(a_{ij}\tilde{\rho}\right)}{\partial x_i\partial x_j},\label{FD2}\\
&&\tilde{\rho}(\zero,x)=\rho_0(x), \quad \tilde{\rho}(T,y)=\rho_T(y).\label{FD3}
\end{eqnarray}
\end{subequations}
The motivation for this specific form of the index comes from a relative entropy problem on path space ({\em Schr\"{o}dinger Bridge Problem}) in the case when $[a_{ij}]$ does not depend on the spatial variable $x$ \cite{F2,W,DPP}. When $[a_{ij}]$ does depend on $x$, such an interpretation is available only under rather restrictive assumptions such as uniform boundness of $a$ \cite[Section 5]{Fischer}. Problem (\ref{FDproblem}) can therefore be viewed as a generalization of the {\mike classical probabilistic }Schr\"{o}dinger bridges problem.

The variational analysis for (\ref{FDproblem}) can be carried out as follows. Let $\mathcal X_{\rho_0\rho_T}$ be the family of flows of probability densities
\[\tilde{\rho}=\{\tilde{\rho}(\cdot,t)\mid \zero\le t\le T\}\]
satisfying (\ref{FD3}). Let $\mathcal U$ be the family of continuous  feedback control laws $u(\cdot,\cdot)$. Consider the unconstrained minimization of the Lagrangian over $\mathcal X_{\rho_0\rho_T}\times\mathcal U$
\begin{eqnarray}\nonumber
\mathcal L(\tilde{\rho},u,\lambda)=\int_{\mR^\N}\int_{\zero}^{T}\left[\left(\frac{1}{2}\|u(x,t)\|^2+V(x,t)\right)\tilde{\rho}(x,t)\right.\\\nonumber
-\lambda(x,t)\left(\frac{\partial \tilde{\rho}}{\partial t}+\nabla\cdot((f+\sigma u)\tilde{\rho})\right.\\\left.\left.-\frac{1}{2}\sum_{i,j=1}^\N\frac{\partial^2}{\partial x_i\partial x_j}\left(a_{ij}(x,t)\tilde{\rho}\right)\right)\right]dtdx,\nonumber
\end{eqnarray}
where $\lambda$ is a $C^1$ Lagrange multiplier. After integration by parts, assuming that limits for $x\rightarrow\infty$ are zero, and observing that the boundary values are constant over $\mathcal X_{\rho_0\rho_T}$, we get the problem
\begin{eqnarray}\nonumber
&&\hspace*{-1cm}\inf_{(\tilde{\rho},u)\in\mathcal X_{\rho_0\rho_T}\times\mathcal U}\int_{\mR^\N}
\int_{\zero}^{T}\left[\frac{1}{2}\|u\|^2+V+\left(\frac{\partial \lambda}{\partial t}\phantom{\sum_{i,j=1}^\N}
\right.\right.
\\
&&\hspace*{-1cm}\label{lagrangian2}\left.\left.
+(f+\sigma u)\cdot\nabla\lambda+\frac{1}{2}\sum_{i,j=1}^\N a_{ij}\frac{\partial^2\lambda}{\partial x_i\partial x_j}\right)\right]\tilde{\rho}(x,t)dtdx
\end{eqnarray}
Pointwise minimization of the integrand with respect to $u$ for each fixed flow of probability densities $\tilde{\rho}$ gives
\begin{equation}\label{prioroptcond}
u^*_{\tilde{\rho}}(x,t)=-\sigma'\nabla\lambda(x,t).
\end{equation}
Plugging this form of the optimal control into (\ref{lagrangian2}), we get the functional of $\tilde{\rho}\in\mathcal X_{\rho_0\rho_T}$
\begin{eqnarray}\label{priorlagrangian3}
J(\tilde{\rho},\lambda)=\int_{\mR^\N}\int_{\zero}^{T}\left[\frac{\partial \lambda}{\partial t}+f\cdot\nabla\lambda-\frac{1}{2}\nabla\lambda\cdot a\nabla\lambda\right.\\
\left.+V+\frac{1}{2}\sum_{i,j=1}^\N a_{ij}(x,t)\frac{\partial^2\lambda}{\partial x_i\partial x_j}\right]\tilde{\rho}(x,t)dtdx.
\end{eqnarray}
We then have the following result:
\begin{prop}If $\tilde{\rho}^*$ satisfies
\begin{equation}\label{optev}
\frac{\partial \tilde{\rho}}{\partial t}+\nabla\cdot((f-a\nabla\lambda)\tilde{\rho})=\frac{1}{2}\sum_{i,j=1}^\N\frac{\partial^2\left(a_{ij}\tilde{\rho}\right)}{\partial x_i\partial x_j},
\end{equation}
with $\lambda$ a solution of the HJB-like equation
\begin{equation}\label{HJB}
\frac{\partial \lambda}{\partial t}+f\cdot\nabla\lambda+\frac{1}{2}\sum_{i,j=1}^\N a_{ij}(x,t)\frac{\partial^2\lambda}{\partial x_i\partial x_j}=\frac{1}{2}\nabla\lambda\cdot a\nabla\lambda-V,
\end{equation}
and $\tilde{\rho}^*(x,T)=\rho_T(x)$, then the pair $\left(\tilde{\rho}^*,u^*\right)$ with $u^*=-\sigma'\nabla\lambda$ is a solution of  (\ref{FDproblem}).
\end{prop}
Of course, the difficulty lies with the nonlinear equation (\ref{HJB}) for which no boundary value is available. Together, $\tilde \rho(x,t)$ and $\lambda(x,t)$ satisfy the coupled equations (\ref{optev}-\ref{HJB}) and the split boundary conditions for $\tilde \rho(x,t)$ in \eqref{FD3}.
However, let us define
$$\varphi(x,t)=\exp[-\lambda(x,t)], \quad (x,t)\in\mR^\N\times [\zero,T].
$$
If $\lambda$ satisfies (\ref{HJB}), we get that $\varphi$ satisfies the {\em linear} equation
\begin{equation}\label{harmonic}
\frac{\partial \varphi}{\partial t}+f\cdot\nabla\varphi+\frac{1}{2}\sum_{i,j=1}^\N a_{ij}(x,t)\frac{\partial^2\varphi}{\partial x_i\partial x_j}=V\varphi
\end{equation}
Moreover, for $\tilde{\rho}$ satisfying (\ref{optev}) and $\varphi$ satisfying (\ref{harmonic}), let us define
$$\hat{\varphi}(x,t)=\frac{\tilde{\rho}(x,t)}{\varphi(x,t)},\quad (x,t)\in\mR^\N\times [\zero,T].
$$
Then a long but straightforward calculation  shows that $\hat{\varphi}$ satisfies the original  equation (\ref{Fokker-Planck}). Thus, we have the system of linear PDE's
\begin{subequations}\label{generalizedschrodinger}\begin{eqnarray}\label{shroedinger1}
\hspace*{-.3in}\frac{\partial \varphi}{\partial t}+f(x,t)\cdot\nabla\varphi+\frac{1}{2}\sum_{i,j=1}^\N a_{ij}\frac{\partial^2\varphi}{\partial x_i\partial x_j}&=&V\varphi,\\
\hspace*{-.3in}\frac{\partial \hat{\varphi}}{\partial t}+\nabla\cdot(f(x,t)\hat{\varphi})-\frac{1}{2}\sum_{i,j=1}^\N\frac{\partial^2\left(a_{ij}\hat{\varphi}\right)}{\partial x_i\partial x_j}&=&-V\hat{\varphi},
\label{schroedinger2}\end{eqnarray}
nonlinearly coupled through their boundary values as
\begin{equation}\label{BND}
\varphi(x,\zero)\hat{\varphi}(x,\zero)=\tilde\rho_0(x),\quad \varphi(x,T)\hat{\varphi}(x,T)=\tilde\rho_T(x).
\end{equation}
\end{subequations}
Equations (\ref{shroedinger1})-(\ref{BND}) constitute a {\em generalized Schr\"{o}dinger system}.  {\mike We have therefore established the following result.
\begin{thm}Let $(\varphi(x,t),\hat{\varphi}(x,t))$ be nonnegative functions satisfying (\ref{shroedinger1})-(\ref{BND}) for $(x,t)\in\left(\mR^n\times [0,T]\right)$. Suppose $\varphi$ is everywhere positive. Then the pair $\left(\tilde{\rho}^*,u^*\right)$ with
\begin{subequations}
\begin{eqnarray}\label{optlaw}
\hspace*{-.3in}u^*(x,t)&=&\sigma'\nabla\log\varphi(x,t),\\
\hspace*{-.3in}\frac{\partial \tilde{\rho}}{\partial t}+\nabla\cdot((f+a\nabla\log\varphi)\tilde{\rho})&=&\frac{1}{2}\sum_{i,j=1}^\N\frac{\partial^2\left(a_{ij}\tilde{\rho}\right)}{\partial x_i\partial x_j},
\label{optevolution}\end{eqnarray}
\end{subequations}
is a solution of  (\ref{FDproblem}).
\end{thm}
Establishing existence and uniqueness (up to multiplication/division of the two functions by a positive constant) of the solution of the Schr\"{o}dinger system is extremely challenging even when the diffusion coefficient matrix $a$ is constant and nonsingular. Nevertheless, if the fundamental solution $p$ of (\ref{Fokker-Planck}) is everywhere positive on $\left(\mR^n\times (0,T]\right)$, existence and uniqueness follows from a deep result of Beurling \cite{beurling} suitably extended by Jamison \cite[Theorem 3.2]{J}, \cite[Section 10]{W}.}

\begin{remark} It is interesting to note that although \eqref{FDproblem} is not convex in $(\tilde\rho,u)$, it can be turned into a convex problem in a new set of coordinates $(\tilde \rho, \tilde m)$ where $m=\tilde\rho  u$, in which case it becomes
\begin{subequations}\label{FDproblem_convex}
\begin{eqnarray}
\label{FD1_convex}
&&\hspace*{-.3in}\inf_{(\tilde{\rho},\tilde{m})}\int_{\mR^\N}\int_{\zero}^{T}\left[\frac{1}{2}\frac{\|m\|^2}{\tilde{\rho}(x,t)}+V(x,t)\tilde{\rho}(x,t)\right]dtdx,\\
&&\hspace*{-.3in}\frac{\partial \tilde{\rho}}{\partial t}+\nabla\cdot(f\tilde{\rho}+\sigma m)=\frac{1}{2}\sum_{i,j=1}^\N\frac{\partial^2\left(a_{ij}\tilde{\rho}\right)}{\partial x_i\partial x_j},\label{FD2_convex}\\
&&\hspace*{-.3in}\tilde{\rho}(\zero,x)=\rho_0(x), \quad \tilde{\rho}(T,y)=\rho_T(y).\label{FD3_convex}
\end{eqnarray}
\end{subequations}
This type of coordinate transformation has been effectively used in \cite{BB} in the context of optimal mass transport.
\end{remark}

\section{The linear-quadratic case}\label{sec:linearS}

We now specialize system \eqref{generalizedschrodinger} to the case of linear dynamics {\mike with constant diffusion matrix} and quadratic loss function $V(x)$, i.e., we assume that $\rho(x,t)$ represents the density function of a linear diffusion
\begin{equation}\label{eq:linearSDE}
dX(t)=AX(t)dt+Bu(t)+Bdw(t), \mbox{ with }X(0)=\xi,\mbox{ a.s.}
\end{equation}
and $\xi$ distributed according to
\begin{equation}\label{initial}\rho(x,0)=\frac{1}{\sqrt{(2\pi)^{n}\det (\Sigma_0)}}\exp\left(-\frac{1}{2}x'\Sigma_0^{-1}x\right)\nonumber
\end{equation}
with $\Sigma_0>0$.
We also assume a loss/state-cost function
\[V(x,t)=\frac12 x' S(t) x\]
and
a ``target'' end-point distribution
\begin{equation}\label{final}\rho(x,T)=\frac{1}{\sqrt{(2\pi)^{n}\det (\Sigma_T)}}\exp\left(-\frac{1}{2}x'\Sigma_T^{-1}x\right),\nonumber
\end{equation}
at $t=T$, with $S(t)\geq 0$ and $\Sigma_T>0$ \footnote{{\mike The special case where the diffusion coefficient $BB'$ is positive definite and $V\equiv 0$ has been studied in \cite{VP}.}} .

\newcommand{\HH}{{\rm H}}
Take $\varphi(x,t)$ and $\hat{\varphi}(x,t)$ in the form
\begin{eqnarray*}
\varphi(x,t)&=&c(t)\exp\{-\frac12 x'\Pi(t)x\}\\
\hat{\varphi}(x,t)&=&\hat c(t)\exp\{-\frac12 x'\HH(t)x\}.
\end{eqnarray*}
Substitution into \eqref{generalizedschrodinger} and separation of variables leads after straightforward calculation to the following two coupled Riccati equations with split boundary conditions
\begin{subequations}\label{LQschrodinger}
\begin{eqnarray}
\label{LQschrodinger1}
\hspace*{-.3in}-\dot\Pi(t)&=&A'\Pi(t)+\Pi(t)A-\Pi(t)BB'\Pi(t) +S(t)\\
\label{LQschrodinger2}
\hspace*{-.3in}-\dot\HH(t)&=&A'\HH(t)+\HH(t)A+\HH(t)BB'\HH(t) -S(t)
\end{eqnarray}
with
\begin{eqnarray}\label{LQschrodinger3}
&&\hspace*{-.3in}\Sigma_0^{-1}=\Pi(0)+\HH(0)
\mbox{ and }\Sigma_T^{-1}=\Pi(T)+\HH(T).
\end{eqnarray}
and
\begin{eqnarray*}
c(t)&=&\exp\{\frac12 \int_0^t{\rm trace}(BB'\Pi(\tau))d\tau\}\\
{\hat c}(t)&=&\exp\left\{- \int_0^t{\rm trace}\left[A(\tau)+\frac12BB'\HH(\tau)\right]d\tau\right\}
\end{eqnarray*}
\end{subequations}
Thus, the problem reduces to finding a pair $(\Pi(t),\HH(t))$ satisfying (\ref{LQschrodinger}). For the case when $S(t)\equiv 0$, it has been shown by the authors in \cite{CGP} that this system has a unique solution.

\section{Semi-definite programming formulation}\label{sec:numerics}

It appears that the solution of (\ref{LQschrodinger1}-\ref{LQschrodinger3}) using successive approximation is not numerically stable. Thus, we now present an alternative formulation into semi-definite program. This allows computation of suboptimal solutions that are arbitrarily close to being optimal. This method was used to work out the example that follows in Section \ref{sec:example}.

We are interested in computing a feedback gain $K(t)$ so that
the control signal $u(t)=-K(t)x(t)$ steers \eqref{eq:linearSDE} from the initial state-covariance $\Sigma_0$ at $t=0$ to the final $\Sigma_T$ at $t=T$. The cost functional to be minimized is
    \begin{eqnarray}\label{eq:functional}
       J&=&\hspace*{-.05in} \E\left\{\int_0^T \left[\frac12\|u\|^2+\frac12 x(t)'S(t)x(t) \right]dt\right\}\\
       &=&\hspace*{-.05in}\frac12\int_0^T \left[\tr(K(t)\Sigma(t)K(t)')+\tr(S(t)\Sigma(t))\right]dt\nonumber
    \end{eqnarray}
    subject to the corresponding differential Lyapunov equation for the state covariance
\begin{equation}\label{eq:diffLyapunov}
\dot\Sigma(t)=(A-BK)\Sigma(t)+\Sigma(t) (A-BK)'+BB'
\end{equation}
satisfying the boundary conditions
\begin{subequations}\begin{equation}\label{eq:boundary}
\Sigma(0)=\Sigma_0, \mbox{ and }\Sigma(T)=\Sigma_T.
\end{equation}

If we replace $K$ by $U(t):=-\Sigma(t)K(t)'$, then
\[
J=\frac12\int_0^T [\tr(U(t)'\Sigma(t)^{-1}U(t))+\tr(S(t)\Sigma(t))]dt
\]
becomes {\em jointly convex} in $U(t)$ and $\Sigma(t)$. On the other hand, the Lyapunov equation \eqref{eq:diffLyapunov} becomes
\begin{equation}\label{eq:diffeqB1U}
\dot\Sigma(t)=A\Sigma(t)+\Sigma(t) A'+BU(t)'+U(t)B'+BB'
\end{equation}
and is now {\em linear} in both $U$ and $\Sigma$. Thus, our optimization problem reduces to the semi-definite program to minimize
\begin{equation}\label{eq:sdp}
 \int_0^T [\tr(Y(t))+\tr(S(t)\Sigma(t))]dt
 \end{equation}
 subject to (\ref{eq:boundary}-\ref{eq:diffeqB1U}) and
 \begin{equation}\label{eq:last}
\left[\begin{matrix}Y(t)& U(t)' \\U(t) & \Sigma(t)\end{matrix}\right]\ge 0.
\end{equation}
\end{subequations}
After discretization in time, (\ref{eq:boundary}-\ref{eq:last}) can be solved numerically and a (suboptimal) gain recovered as
\[K(t)=-U(t)'\Sigma(t)^{-1}.\]

\section{Example}\label{sec:example}
We consider inertial particles modeled by
   \begin{eqnarray*}
       dx(t) &=& v(t)dt \\
       dv(t) &=& u(t)dt+ dw(t).
   \end{eqnarray*}
Here, $u(t)$ is a control input (force) at our disposal, $x(t)$ represents the position and $v(t)$ velocity of particles, while $w(t)$ represents random exitation (corresponding to ``white noise'' forcing).
We wish to steer the spread of the particles from an initial Gaussian distribution with $\Sigma_0=2I$ at $t=0$ to the terminal marginal $\Sigma_T=1/4I$
for $T=1$ in a optimal way such that the cost function \eqref{eq:functional} is minimized.

\begin{figure}\begin{center}
\includegraphics[width=0.47\textwidth]{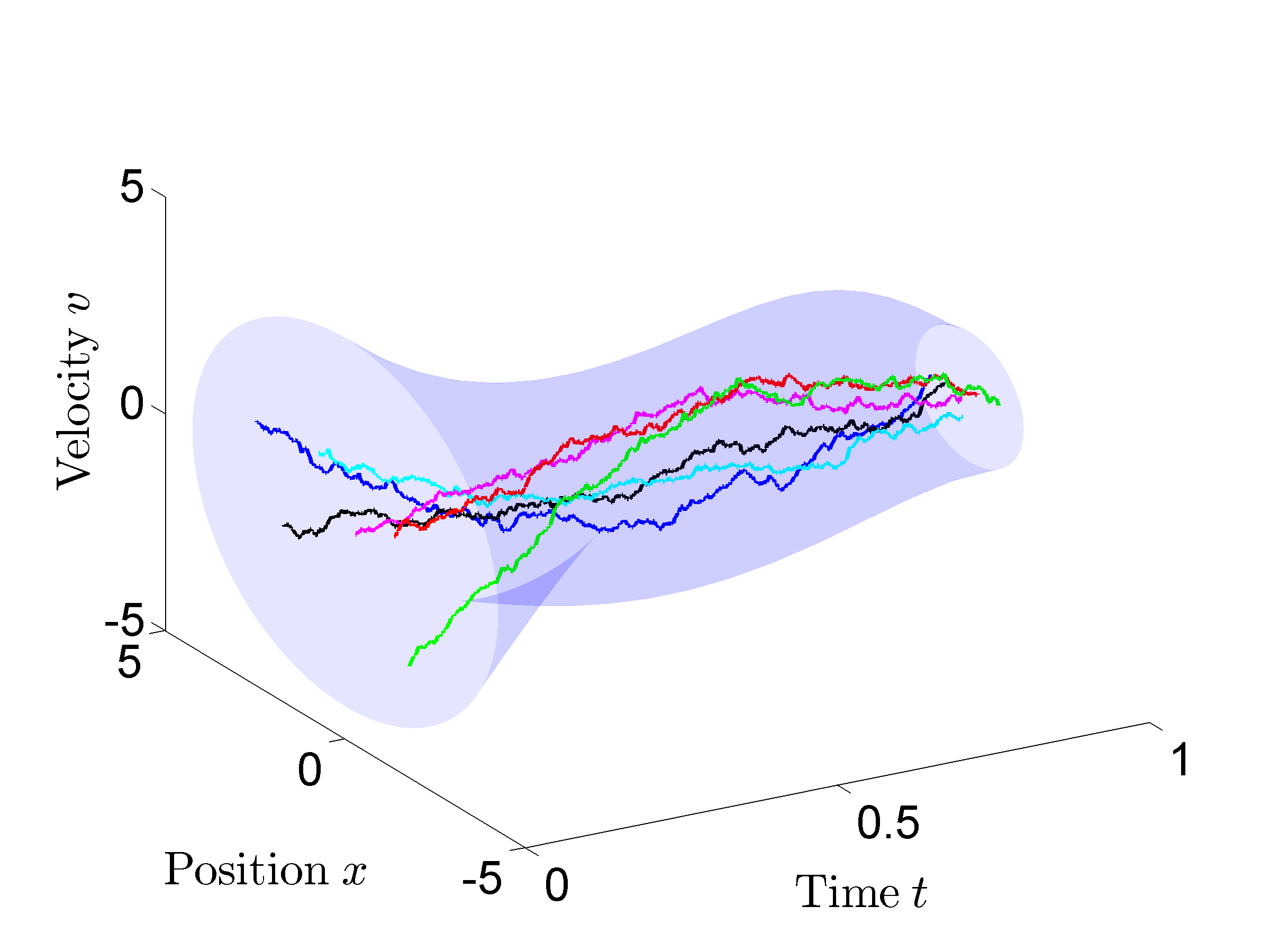}
   \caption{Inertial particles: state trajectories ($S(t)\equiv I$)}
   \label{fig:Eg1Phase1}
\end{center}\end{figure}

\begin{figure}\begin{center}
\includegraphics[width=0.47\textwidth]{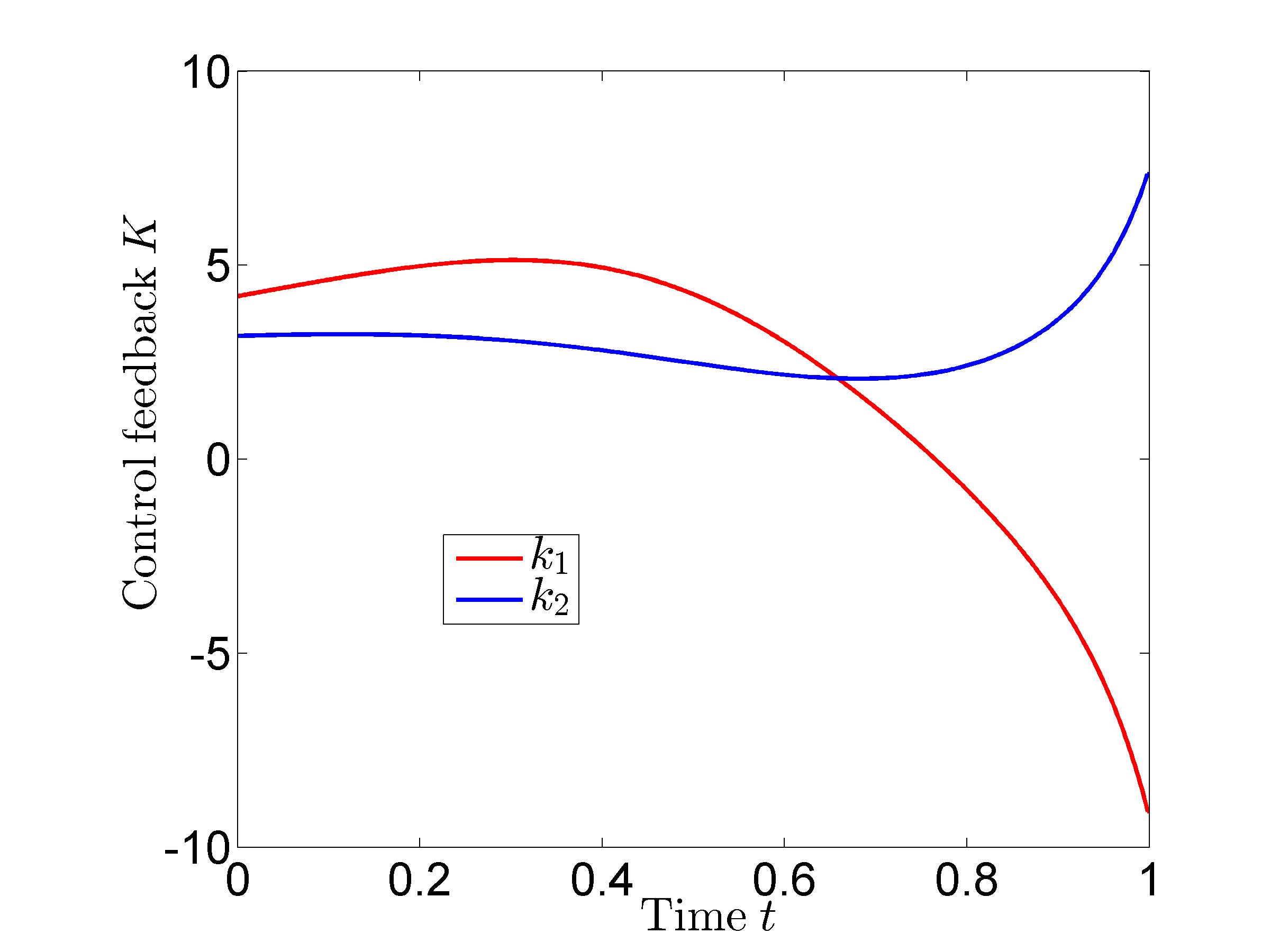}
   \caption{Inertial particles: feedback gains}
   \label{fig:Eg1Controlfeedback}
\end{center}\end{figure}

\begin{figure}\begin{center}
\includegraphics[width=0.47\textwidth]{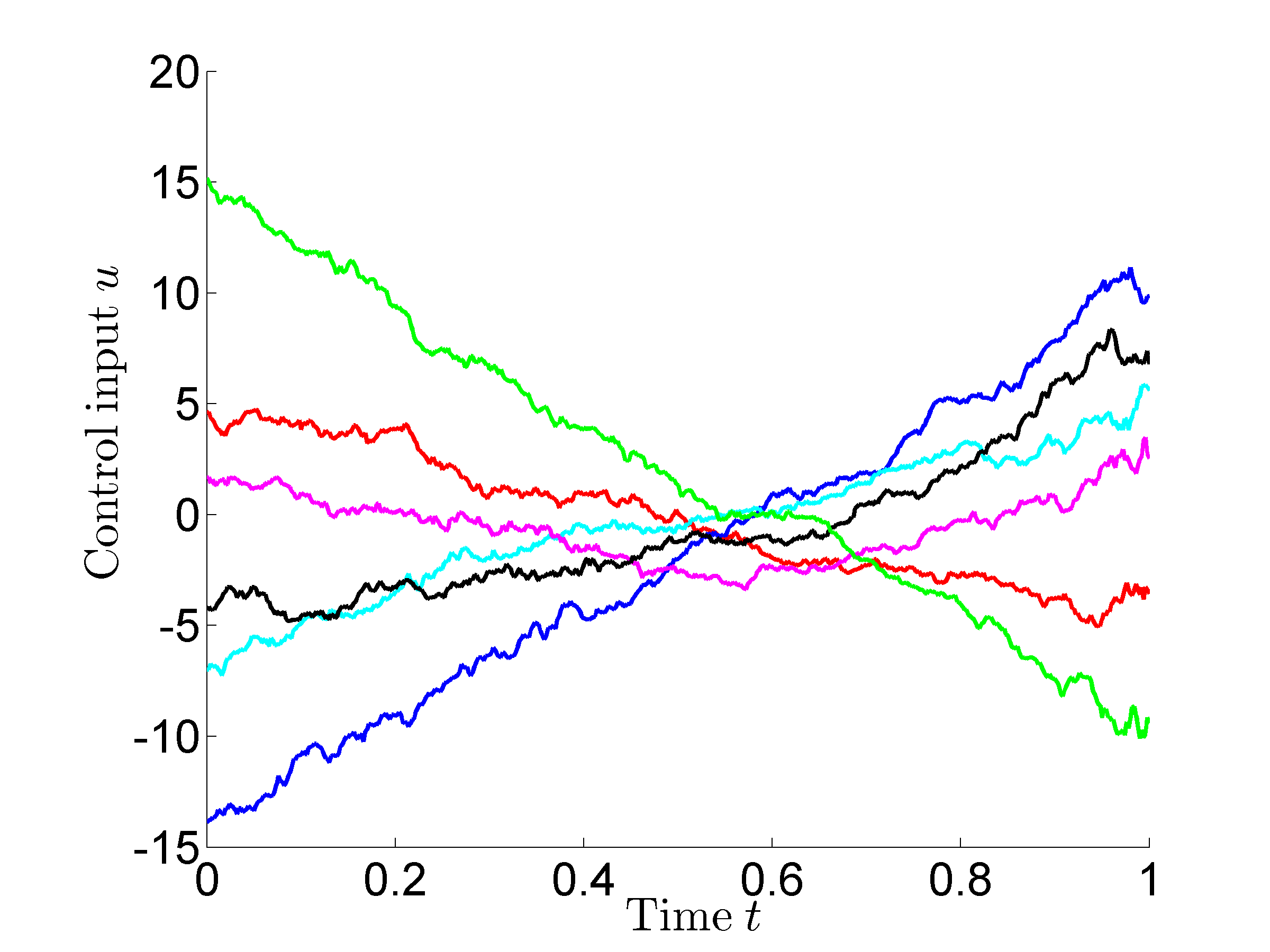}
   \caption{Inertial particles: control inputs}
   \label{fig:Eg1Control1}
\end{center}\end{figure}

Figure~\ref{fig:Eg1Phase1} displays typical sample paths $\{(x(t),v(t))\mid t\in[0,1]\}$ in phase space, as a function of time, that are attained using the optimal feedback strategy derived following \eqref{eq:sdp} and $S=I$.
The feedback gains $K(t)=[k_1(t),\,k_2(t)]$ are shown in Figure \ref{fig:Eg1Controlfeedback} as a function of time.
Figure \ref{fig:Eg1Control1} shows the corresponding control action for each trajectory.

\begin{figure}\begin{center}
\includegraphics[width=0.47\textwidth]{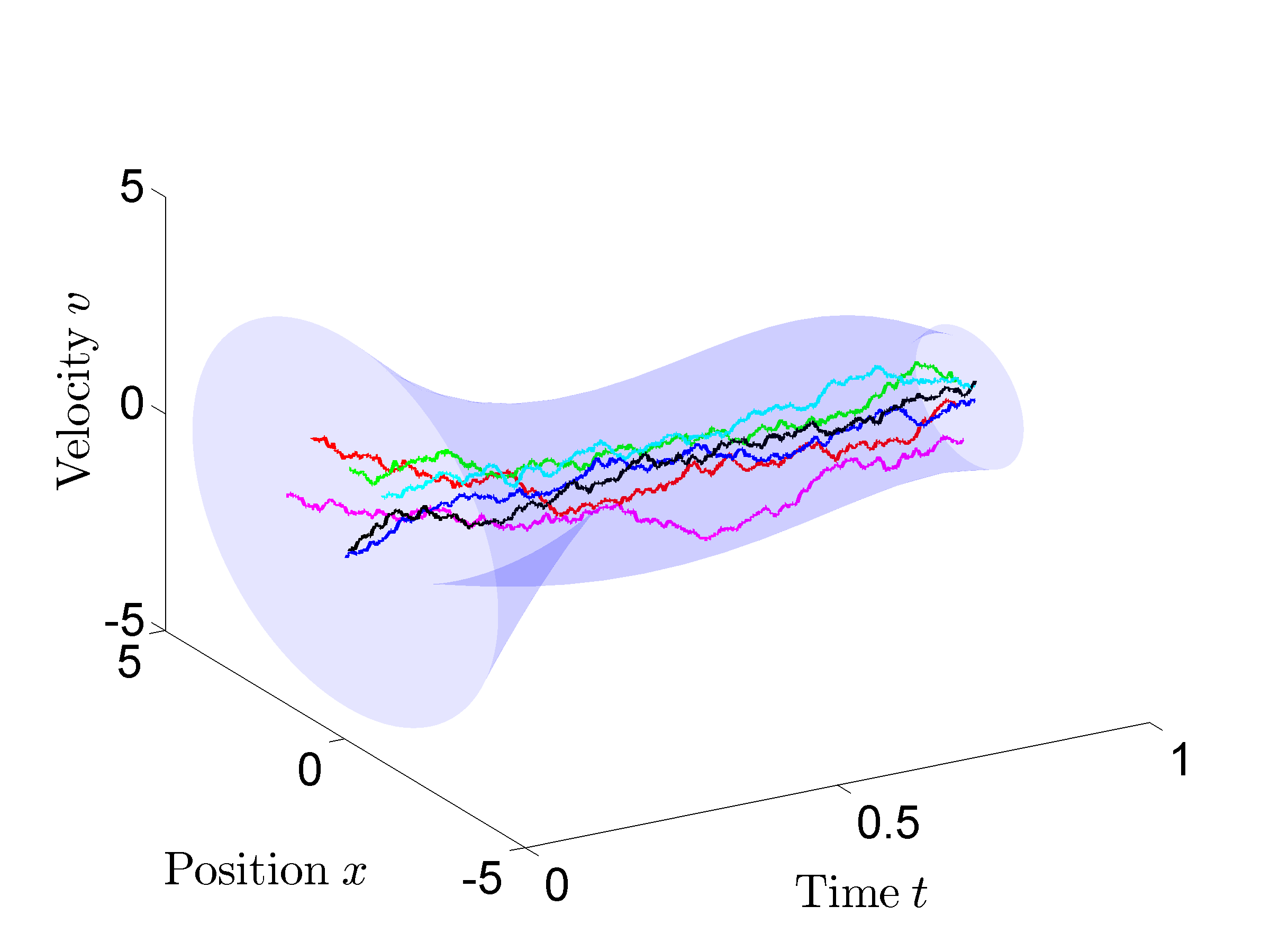}
   \caption{Inertial particles: state trajectories ($S(t)\equiv10I$)}
   \label{fig:Eg1Phase2}
\end{center}\end{figure}

For comparison, Figure \ref{fig:Eg1Phase2} displays typical sample paths when optimal control is used and $S=10I$. As expected, $\Sigma(\cdot)$ shrinks faster as we increase the state penalty $S$ {\mike which is consistent with the reference evolution loosing probability mass at a higher rate at places where $V(x)$ is large. }

\spacingset{.97}
\bibliographystyle{IEEEtran}
\bibliography{refs}

\begin{thebibliography}{10}
\providecommand{\url}[1]{#1}
\csname url@samestyle\endcsname
\providecommand{\newblock}{\relax}
\providecommand{\bibinfo}[2]{#2}
\providecommand{\BIBentrySTDinterwordspacing}{\spaceskip=0pt\relax}
\providecommand{\BIBentryALTinterwordstretchfactor}{4}
\providecommand{\BIBentryALTinterwordspacing}{\spaceskip=\fontdimen2\font plus
\BIBentryALTinterwordstretchfactor\fontdimen3\font minus
  \fontdimen4\font\relax}
\providecommand{\BIBforeignlanguage}[2]{{%
\expandafter\ifx\csname l@#1\endcsname\relax
\typeout{** WARNING: IEEEtran.bst: No hyphenation pattern has been}%
\typeout{** loaded for the language `#1'. Using the pattern for}%
\typeout{** the default language instead.}%
\else
\language=\csname l@#1\endcsname
\fi
#2}}
\providecommand{\BIBdecl}{\relax}
\BIBdecl

\bibitem{Schrodinger1}
E.~Schr{\"o}dinger, \emph{{\"U}ber die umkehrung der naturgesetze}.\hskip 1em
  plus 0.5em minus 0.4em\relax Verlag Akademie der wissenschaften in kommission
  bei Walter de Gruyter u. Company, 1931.

\bibitem{Schrodinger2}
------, ``Sur la th{\'e}orie relativiste de l'{\'e}lectron et
  l'interpr{\'e}tation de la m{\'e}canique quantique,'' in \emph{Annales de
  l'institut Henri Poincar{\'e}}, vol.~2, no.~4.\hskip 1em plus 0.5em minus
  0.4em\relax Presses universitaires de France, 1932, pp. 269--310.

\bibitem{W}
A.~Wakolbinger, ``Schr{\"o}dinger bridges from 1931 to 1991,'' in \emph{Proc.
  of the 4th Latin American Congress in Probability and Mathematical
  Statistics, Mexico City}, 1990, pp. 61--79.

\bibitem{F2}
H.~F{\"o}llmer, ``Random fields and diffusion processes,'' in \emph{{\'E}cole
  d'{\'E}t{\'e} de Probabilit{\'e}s de Saint-Flour XV--XVII, 1985--87}.\hskip
  1em plus 0.5em minus 0.4em\relax Springer, 1988, pp. 101--203.

\bibitem{FSA}
S.~Feghhi, M.~Shahriari, and H.~Afarideh, ``Calculation of neutron importance
  function in fissionable assemblies using {M}onte {C}arlo method,''
  \emph{Annals of Nuclear Energy}, vol.~34, no.~6, pp. 514--520, 2007.

\bibitem{DP}
P.~Dai~Pra, ``A stochastic control approach to reciprocal diffusion
  processes,'' \emph{Applied mathematics and Optimization}, vol.~23, no.~1, pp.
  313--329, 1991.

\bibitem{PW}
M.~Pavon and A.~Wakolbinger, ``On free energy, stochastic control, and
  {S}chr{\"o}dinger processes,'' in \emph{Modeling, Estimation and Control of
  Systems with Uncertainty}.\hskip 1em plus 0.5em minus 0.4em\relax Springer,
  1991, pp. 334--348.

\bibitem{DGW}
D.~Dawson, L.~Gorostiza, and A.~Wakolbinger, ``Schr{\"o}dinger processes and
  large deviations,'' \emph{Journal of mathematical physics}, vol.~31, no.~10,
  pp. 2385--2388, 1990.

\bibitem{DPP}
P.~Dai~Pra and M.~Pavon, ``On the {M}arkov processes of {S}chr{\"o}dinger, the
  {F}eynman-{K}ac formula and stochastic control,'' in \emph{Realization and
  Modelling in System Theory}.\hskip 1em plus 0.5em minus 0.4em\relax Springer,
  1990, pp. 497--504.

\bibitem{Vinante}
A.~Vinante, M.~Bignotto, M.~Bonaldi, M.~Cerdonio, L.~Conti, P.~Falferi,
  N.~Liguori, S.~Longo, R.~Mezzena, A.~Ortolan \emph{et~al.}, ``Feedback
  cooling of the normal modes of a massive electromechanical system to
  submillikelvin temperature,'' \emph{Physical review letters}, vol. 101,
  no.~3, p. 033601, 2008.

\bibitem{Munakata}
T.~Munakata and M.~Rosinberg, ``Feedback cooling, measurement errors, and
  entropy production,'' \emph{Journal of Statistical Mechanics: Theory and
  Experiment}, vol. 2013, no.~06, p. P06014, 2013.

\bibitem{CGP}
Y.~Chen, T.~Georgiou, and M.~Pavon, ``Optimal steering of a linear stochastic
  system to a final probability distribution,'' \emph{arXiv preprint
  arXiv:1408.2222}, 2014.

\bibitem{CGP2}
------, ``Optimal steering of a linear stochastic system to a final probability
  distribution, part {II},'' \emph{in preparation}.

\bibitem{CGP3}
------, ``Fast cooling for a system of stochastic oscillators,'' \emph{in
  preparation}.

\bibitem{PM}
P.~Perona and J.~Malik, ``Scale-space and edge detection using anisotropic
  diffusion,'' \emph{Pattern Analysis and Machine Intelligence, IEEE
  Transactions on}, vol.~12, no.~7, pp. 629--639, 1990.

\bibitem{G}
P.~Guidotti, ``Anisotropic diffusions of image processing from {P}erona-{M}alik
  on,'' \emph{Advanced Studies in Pure Mathematics}, vol.~99, p. 20XX, 2007.

\bibitem{Fischer}
M.~Fischer, ``On the form of the large deviation rate function for the
  empirical measures of weakly interacting systems,'' \emph{Bernoulli},
  vol.~20, no.~4, pp. 1765--1801, 2014.

\bibitem{KS}
I.~Karatzas, \emph{Brownian motion and stochastic calculus}.\hskip 1em plus
  0.5em minus 0.4em\relax springer, 1991, vol. 113.

\bibitem{BB}
J.-D. Benamou and Y.~Brenier, ``A computational fluid mechanics solution to the
  {M}onge-{K}antorovich mass transfer problem,'' \emph{Numerische Mathematik},
  vol.~84, no.~3, pp. 375--393, 2000.

\bibitem{Vil}
C.~Villani, \emph{Topics in optimal transportation}.\hskip 1em plus 0.5em minus
  0.4em\relax American Mathematical Soc., 2003, no.~58.

\bibitem{AGS}
L.~Ambrosio, N.~Gigli, and G.~Savar{\'e}, \emph{Gradient flows: in metric
  spaces and in the space of probability measures}.\hskip 1em plus 0.5em minus
  0.4em\relax Springer, 2006.

\bibitem{KT}
S.~Karlin and H.~M. Taylor, \emph{A second course in stochastic
  processes}.\hskip 1em plus 0.5em minus 0.4em\relax Gulf Professional
  Publishing, 1981, vol.~2.

\bibitem{H}
L.~H{\"o}rmander, ``Hypoelliptic second order differential equations,''
  \emph{Acta Mathematica}, vol. 119, no.~1, pp. 147--171, 1967.

\bibitem{OR}
O.~A. Oleinik and E.~V. Radkevich, ``Second order equations with nonnegative
  characteristic form,'' \emph{Itogi Nauki i Tekhniki. Seriya" Matematicheskii
  Analiz"}, vol.~7, pp. 7--252, 1971.

\bibitem{BCGR}
U.~Boscain, R.~Chertovskih, J.-P. Gauthier, and A.~Remizov, ``Hypoelliptic
  diffusion and human vision: A semidiscrete new twist,'' \emph{SIAM Journal on
  Imaging Sciences}, vol.~7, no.~2, pp. 669--695, 2014.

\bibitem{beurling}
A.~Beurling, ``An automorphism of product measures,'' \emph{The Annals of
  Mathematics}, vol.~72, no.~1, pp. 189--200, 1960.

\bibitem{J}
B.~Jamison, ``Reciprocal processes,'' \emph{Z. Wahrscheinlichkeitstheorie verw.
  Gebiete}, vol.~30, pp. 65--86, 1974.

\bibitem{VP}
I.~G. Vladimirov and I.~R. Petersen, ``Minimum relative entropy state
  transitions in linear stochastic systems: the continuous time case,'' in
  \emph{Proceedings of 19th International Symposium on Mathematical Theory of
  Networks and Systems}, 2010, pp. 51--58.

\end{thebibliography}
\end{document}